\documentclass[pre,twocolumn]{revtex4}
\usepackage{psfrag}
\usepackage{amssymb,amsmath,amsthm}
\usepackage[pdftex]{graphicx}
\usepackage{verbatim}
\newcommand{\bm}[1]{\mbox{\boldmath$#1$}}

\begin{document}
\title{INFLUENCE OF SEDIMENTATION ON CONVECTIVE INSTABILITIES IN COLLOIDAL SUSPENSIONS}
\author{ANDREY RYSKIN* and HARALD PLEINER
    }
\affiliation{Max Planck Institute for Polymer Research, Mainz, Germany}

\begin{abstract} \noindent
We investigate theoretically the bifurcation scenario for colloidal suspensions subject to a vertical temperature gradient taking into account the effect of sedimentation. In contrast to molecular binary mixtures, here the thermal relaxation time is much shorter than that for concentration fluctuations.  This allows for differently prepared ground states, where a concentration profile due to sedimentation and/or the Soret effect has been established or not. This gives rise to different linear instability behaviors, which are manifest in the temporal evolution into the final, generally stationary convective state. In a certain range above a rather high barometric number there is a coexistence between the quiescent state and the stationary convective one, allowing for a hysteretic scenario.   
\noindent 

\bigskip


\end{abstract}

 \maketitle

\noindent {\bf 1. Introduction} \smallskip

\noindent Thermal convection in binary mixtures has
attracted much research activity already for  along time [Platten \& Legros 1984, Cross \& Hohenberg 1993, L\"ucke et al. 1998]. In comparison to the
pure fluid case, the dynamics and the bifurcation scenario are
more complicated due to the extra degree of freedom associated
with the concentration field. Thereby solutal currents are not
only driven by concentration gradients, they occur also in
response to temperature inhomogeneities. This is denoted as the
thermo-diffusive or Soret effect. Its influence on the convective
buoyancy force is quantified by the dimensionless separation
ratio $\psi$. The sign of $\psi$ indicates whether temperature-
and solutal-induced density gradients are parallel or
opposed to each other. At negative $\psi$ the motionless
conductive state experiences an oscillatory instability,
saturating  in a nonlinear state of traveling waves
[L\"ucke et al. 1998]. On the other hand, at positive $\psi$ the
convective instability remains stationary, but the critical
Rayleigh number for the onset of convection is dramatically
reduced as compared to the pure-fluid reference value
$Ra_c^0=1708$. This is a result of the joint action of thermal and
solutal buoyancy forces.

A typical property of binary mixture convection is the formation
of  concentration boundary layers [Winkler \& Kolodner 1992]. This is a
consequence of the fact that  the concentration diffusivity $D_c$
in mixtures is usually much smaller than the heat diffusivity
$\kappa$.  For molecular binary mixtures the dimensionless Lewis
number $L=D_c/\kappa$ adopts  typical values between $0.1$ and
$0.01$ [Kolodner 1988]. If colloidal suspensions are under
consideration, the time scale separation is even more dramatic. In
this context magnetic colloids, known as ferrofluids, are a
canonical example. These materials are dispersions of heavy solid
ferromagnetic grains suspended in a carrier liquid
[Rosensweig 1985]. With a typical diameter of $10$ nm the
particles are pretty large on molecular length scales, resulting
in an extremely small particle mobility. This feature is reflected
by Lewis numbers as small as $L=10^{-4}$ [Blums et al. 1997]. The
smallness of $L$ leads to a situation where de-mixing effects
take place on very large time scales. Thus, in those experiments, where thermodiffusion is irrelevant,
ferrofluids can safely be treated as single-component fluid systems.

However, ferrofluids and other colloidal suspensions are also known to exhibit a very large
separation ratio $\psi$ (up to $|\psi| \approx 100$ [Blums et al. 1999, Lenglet et al. 2002]. This observation is due to the pronounced thermo-diffusivity of these materials in combination with the large specific weight difference of the two constituents. 
As a result, in these materials the solutal buoyancy forces are rather strong and a two-component treatment of convective instabilities is mandatory. By considering the classical Rayleigh B\'{e}nard setup it is shown [Ryskin et al. 2003] that the convective behavior is significantly different from the case of molecular mixtures. Starting from the motionless configuration with an initially uniform concentration distribution, convective perturbations are found to grow even at Rayleigh numbers well below the threshold $Ra^{0}$ of pure-fluid convection. The actual critical Rayleigh number $Ra_c$ is drastically smaller, but experimentally inaccessible due to the extremely slow growth of convection patterns for $Ra \gtrsim Ra_{c}$, requiring very large observation times. On the other hand, operating the colloidal convection experiment at Rayleigh numbers $Ra_c < Ra \stackrel{<}{\sim} Ra^0$, reveals considerable positive  growth rates, which lead to a saturated nonlinear state almost as fast as pure-fluid convection does at $Ra>Ra^0$.

In an external magnetic field the apparent imperfection of the bifurcation is even more pronounced in the case of ferrofluids. Magnetophoretic effects as well as magnetic stresses have been taken into account in the static and dynamic parts of the equations leading to rather pronounced boundary layer profiles (with respect to the concentration and magnetic potential). This boundary layer couples effectively to the bulk behavior due to the magnetic boundary condition [Ryskin \& Pleiner 2004]. 

In the case of a negative separation ratio (negative Soret coefficient) the thermal and solutal density gradients are opposed when heating from below. The linear convective oscillatory instability known from molecular binary mixtures (with $\psi <-1$) at $Ra^0$ is also found for colloidal ones, but the nonlinear treatment shows that the linearly unstable oscillatory states are  transients only and decay after some time, rendering the final convection-free state stable [Ryskin \& Pleiner 2005]. Above a second threshold, somewhat higher than $Ra^0$, a finite amplitude stationary instability is found, while small amplitude disturbances do not destroy the convection-free state. The traveling wave solution dominating in molecular binary mixtures is shifted to unrealistically high temperature gradients and is not possible anymore in colloidal systems [Huke et al. 2000, Huke et al. 2007]. When heating from above molecular binary mixtures with a negative separation ratio $\psi <-1$,  a linear stationary instability is found, which is basically driven by the solutal buoyancy and only slightly modified by thermal variations. In colloidal suspensions, however, the concentration and temperature dynamics show completely different behavior. Thus, this stationary instability is very different from that obtained by heating from below with a positive separation ratio. In the former case small scale structures arise at very high Ra numbers, whose wavelength decreases strongly with increasing $Ra$. 

In earth's (vertical) gravity field, the density contrast in colloidal suspensions results in a tendency to phase separate the two constituents. However, for truly colloidal systems the particles are small enough that Brownian motion successfully prohibits a real phase separation allowing for the binary mixture description. (In the different case of micrometer-sized particles, e.g. magnetorheological fluids, a two-fluid description [Onuki 1989, Milner 1989, 1993, Pleiner \& Harden 2003] should be used.) Nevertheless, there is a slight accumulation of the heavier constituent towards the bottom, i.e. a sedimentation induced concentration gradient, that can be expected to be relevant in systems with a large separation ratio $\psi$, while in molecular binary mixtures this effect generally is negligible.   
Some experiments show that sedimentation strongly affects the qualitative behavior of thermal instabilities in ferrofluids (Bozhko \& Putin [2003], Bozhko et al. [2006], Tynj{\"a}l{\"a} et al. [2006]). A first theoretical discussion has been provided by Shliomis \& Smorodin [2005]. 

In ferrofluids instead of the temperature gradient an external magnetic field produces a destabilizing force (the Kelvin force), which is larger in areas with a higher concentration of magnetic particles. Concentration fluctuations are therefore amplified and can lead to an
instability. Recently, Ryskin \& Pleiner [2007] have shown theoretically that a gravity-stratified ferrofluid indeed becomes convectively unstable in a sufficiently strong external vertical magnetic field. The amplitude of the velocity field is a rather complicated function of time. Initially the amplitudes grow
exponentially with the linear growth rate. For all realistic parameter values the final flow
at long times is stationary. For intermediate times the amplitude saturates at a high value and then decreases considerably to its asymptotic value. The transition between the intermediate high peak value and
the very small, saturated one is due to the fact that the convective
flow effectively reduces the concentration gradient, thus reducing the very basis for the instability. The final stationary state is reached, when the process of building up the concentration gradient due to sedimentation is
balanced by its destruction due to advection. Since the former process
is very slow, only a very small velocity is necessary. In contrast to the velocities the amplitudes of the concentration variations are not small. The stationary concentration profile is essentially neither linear nor constant due to the nonlinear part of the Kelvin force. This is in marked contrast to the thermal convection problem in non-magnetic colloidal suspension.

In this paper we consider a horizontally infinite slab of colloidal suspension (thickness $h$) subject to gravity and a vertical temperature gradient (no magnetic field). We restrict ourselves to the case of a positive separation ratio and heating from below. We provide a comprehensive discussion of the influence of sedimentation on the bifurcation scenario for that case. Sedimentation affects convective instabilities in two stages. First, on the level of possible ground states and their linear stability, and second on the long-time nonlinear asymptotic . Due to the slow particle diffusion one can prepare different ground states. Applying the temperature gradient to a suspension right after its preparation, sedimentation has no time to develop ("homogeneous state") and the linear instability behavior is the same as in the case without sedimentation [Ryskin et al. 2003]. However, waiting long enough until the sedimentation has taken place ("stratified case") and applying then the temperature gradient, leads to situation similar to that of a negative $\psi$ material (and disregarding sedimentation): the destabilizing temperature gradient has to overcome the stabilizing sedimentation-induced concentration gradient, which leads to an enhanced threshold and an oscillatory linear instability. At an even stronger temperature gradient the oscillation frequency goes to zero rendering the linear instability to be stationary. (This latter feature has no counterpart in the case of negative $\psi$ without sedimentation.) After having set up the  mathematical framework in Sec. 2, the linear instabilities will be discussed in more detail in Sec. 3. Here we also have included the experimentally hardly realizable case that the concentration gradient, due to sedimentation as well as due to the thermodiffusion effect, is fully developed ("fully developed state"). All these linear cases influence the nonlinear bifurcation scenario and are manifest as transients. In Sec. 4 we give an approximate analytical solution for the stationary non-linear problem. We predict a hysteretic behavior due to the effects of sedimentation in a parameter range, where the non-convective state coexists with finite amplitude stationary convection state. Numerically we describe the transients to the stationary state, find an oscillatory solution, and give analytical conditions, for which such a solution can exist. These main results are summarized in Sec. 5.

\bigskip

\noindent {\bf 2. Basic Equations} \smallskip

\noindent We consider a slab of colloidal liquid subject to a positive temperature gradient (in z direction) and gravity (in -z direction). The system of equations is the same as for molecular binary mixtures including incompressibility, as well as momentum, heat, and mass conservation reading in Boussinesq approximation [Boussinesq 1903, Platten \& Chavepeyer 1976, Brand et al. 1984] 
\begin{eqnarray}
\bm{\nabla} \cdot \bm{v}&=&0, \label{contin}\\
\partial_t \bm{v} + \bm{v} \cdot \bm{\nabla} \bm{v}&=&-\bm{\nabla} W
+ Pr \, \bm{\nabla}^2 \bm{v} \nonumber\\
&&+ Pr
\, Ra \left [ (T- {\bar T})- \psi (C- {\bar C}) \right ] \bm{e}_z \quad \quad \label{ns}\\
\partial_t T + \bm{v} \cdot \bm{\nabla} T&=&\bm{\nabla}^2 T, \label{temp}\\
\partial_t C + \bm{v} \cdot \bm{\nabla} C &=& L (\bm{\nabla}^2 C + \bm{\nabla}^2
T). \label{conz}
\end{eqnarray}
The relevant variables are the flow velocity $\bm{v}$, temperature $T$, and concentration (of the particles) $C$. The material is characterized by the transport coefficients $\kappa$, $D_c$, $D_s$, and $\nu$ for heat and
particle diffusion, thermodiffusion, and viscosity, respectively. As usual they are used to make the equations dimensionless scaling length with the layer thickness $h$, time with $h^2/\kappa$,
temperature with $\Delta T$, the applied temperature difference, and concentration with $(D_s/D_c) \Delta T$. The quantities
${\bar T}$ and ${\bar C}$ are reference values defined as the
mean values for temperature and concentration. The pressure $W$ is a kind of Lagrange multiplier  that serves to guarantee incompressibility for all times. We are left with three dimensionless numbers governing the bulk material properties,
the Prandtl number $Pr=\nu/\kappa$, the Lewis
number $L=D_c/\kappa$ , and the separation ratio $\psi=D_s \beta_c/(D_c \beta_T)$, where
$\beta_T=-(1/\rho)\partial \rho/\partial T$ and $\beta_c=(1/\rho)
\partial \rho/\partial c$ are the thermal and solutal expansion coefficients.
The dimensionless Rayleigh number $Ra=\beta_T g h^3 \Delta T
/(\kappa \nu)$, with $g$ the earth's gravity constant, is the control parameter for the bifurcation behavior.

Gravity is not only responsible for the driving force, but also for an inhomogeneous distribution of the particles. In equilibrium, the balance between gravity and Brownian motion leads to a Boltzmann distribution for the concentration [Biben et al., 1993] with the sedimentation length $h_s = k_B T/(\chi_T m_p g)$, where $m_p$ is the effective buoyant mass of a particle, and $\chi_T$ is the osmotic compressibility. Since usually $h_s \gg h$, the exponential distribution reduces to a linear concentration profile due to sedimentation
\begin{equation}
C(z) = \overline{C} (1 - z/h_s), \label{profile}
\end{equation}
where $\overline{C}$ is the mean mass fraction of the colloidal particles. To reach this true equilibrium state one has to wait for a rather long time allowing experiments to be started from either this state or from the homogeneous state $C=\bar C$. The transition from the former to the latter state is due to a mass flux of particles, which can be written in the simplest (and dimensional) form as [Blums 2002]
\begin{equation}
{\bm j}_s = -\frac{D_c}{h_s} C \,{\bm e}_z , \label{1}
\end{equation}
This sedimentation current should be added to the diffusive and thermo-diffusive concentration currents in the concentration dynamics Eq.(\ref{conz}), but is generally neglected there by putting $C=\bar C$ resulting in $\bm{\nabla \cdot j}_s =0$. At the rigid (and impermeable) boundaries, however, the sedimentation current cannot be neglected w.r.t. the other concentration currents, since there the total
concentration current has to vanish, which is guaranteed by the (dimensionless) boundary conditions
\begin{eqnarray}
(\partial_z C + \partial_z T)|_{z=\pm 1/2}&=&-\frac{B}{\psi  \,Ra }, \label{bkc}\\
\bm{v}|_{z=\pm 1/2}&=&0, \label{bkv} \\
T|_{z=\pm 1/2}&=&{\bar T} \mp
\frac{1}{2}. \label{bkt} 
\end{eqnarray}
at the upper ($z=1/2$) and the lower ($z=-1/2$) plate. The additional standard boundary conditions for flow and temperature reflect the no-slip condition and the externally applied temperature gradient across the mean temperature $\bar T$. The boundary condition for the concentration variable contains the barometric number $B$ [Shliomis \& Smorodin 2005]
\begin{equation}
B  =\frac{\beta_c\, g \,\overline{C}\, h^4}{\kappa\, \nu\, h_s},
\end{equation}
which can vary considerably, typically from $1$ to $10^5$, due to the strong $h$-dependence. This parameter measures the importance of sedimentation relative to viscosity and thermal diffusion. A strong impact of sedimentation on the bifurcation scenario can be expected for $B \ge \psi\,Ra^0$. For ferrofluids, using a magnetic field $H_0$ instead of the temperature gradient, a magnetic barometric number
$B_m = \chi_c^2\, \overline{C}^2 H_0^2\,h^4 / (\rho\, \bar \epsilon \,\nu^2
h_s^2)$ governs the sedimentation effects on convective instabilities [Ryskin \& Pleiner 2007]. Here, $\chi_c=\partial \chi/\partial C$ describes the concentration dependence of the magnetic susceptibility $\chi$ giving rise to magnetophoresis, and $\bar \epsilon$ is the effective magnetic permeability of the material.

\bigskip

\noindent {\bf 3. Linear Instabilities} \smallskip

\noindent Starting from the homogeneous state (no sedimentation gradient) the linear development of a convective instability is unaffected by sedimentation effects, since there is no concentration gradient to be advected. Thus, the true threshold, $Ra_c $ is by a factor of $\psi$ smaller than in the single-component case, $Ra_c^0$, but the time evolution is too slow for the instability to be observed. At higher $Ra$ numbers the growth rates are sufficiently high, but the system is already in the instability regime mimicking an imperfect bifurcation to a stationary state. 

In the stratified state (with a sedimentation-induced concentration gradient) the instability threshold is higher than in the case before, since the concentration gradient opposes the temperature gradient. Using the analogy to the case of binary mixtures with a negative separation ratio one can expect a linear oscillatory instability above $Ra_c^0$. However, in contrast to the thermo-diffusive concentration gradient, the sedimentation-induced  one is not proportional to the temperature gradient. Thus if the barometric number is too small, sedimentation cannot compete with the temperature gradient and a stationary instability has to expected. Indeed a linear stability analysis along the lines of Ryskin \& Pleiner [2005] gives the threshold condition
\begin{equation}
3 Ra\,  Pr \,\frac{\lambda + 2 \pi^2 L \psi}{\lambda+2 \pi^2} = 3 B Pr + 27 \pi^2 Pr  \lambda  +7
\lambda^2 \label{lam}
\end{equation}
with one additional sedimentation contribution ($\sim B$). Here, the growth rate $\lambda=0$ and $= i \omega$ for the stationary and oscillatory instability, respectively. The former occurs for small barometric numbers $B < B_1 \equiv L \psi Ra^0$ at the threshold
\begin{eqnarray}
Ra_c^{st} = \frac{B}{L\, \psi} \label{stat}
\end{eqnarray}
independent of any wave number, while for $B>B_1$ an oscillatory linear instability is found with 
\begin{eqnarray}
Ra_c^{os} &=& Ra^0 + \frac{27 {Pr}}{14 + 27 Pr} B \label{rac} \\
\omega^2_c &=& \frac{6 Pr}{14 + 27 Pr} \left({B} - B_1 \right) \label{omegac}
\end{eqnarray}
where the critical wave number approximately by $k_c = \pi$ and $Ra^0 = 18 \pi^4$. For typical material parameters, $\psi \sim 10$ and $L\sim 10^{-4}$, the transition occurs at $B_1 \sim 1$.  Equations\@ (\ref{rac}) and (\ref{omegac}) were (for $\psi = 0$) also obtained by Shliomis \& Smorodin [2005], with slightly different numerical factors due to a different choice of trial functions. 

Increasing the $Ra$ number beyond $Ra_c^{os}$ the oscillating frequency $\omega$ starts to decrease, until it vanishes at a certain $Ra_2$, with the result that for $Ra > Ra_2$ the linear instability is stationary again. $Ra_2$ is a complicated function of $B$ with $Ra_2(B_1) = Ra_c^{os} \approx Ra_c^{st}(B_1)$. This disappearance of the oscillation frequency does not have an analog in the case of negative-$\psi$ colloids (without sedimentation) and again, the reason is the independence of the sedimentation current from the temperature gradient. 

The fully developed state with a concentration gradient due to both, sedimentation and thermodiffusion is difficult to realize in experiments. Not only one has to wait initially for  the sedimentation gradient to develop, also after each temperature gradient step one has to wait for the thermo-diffusive response to be finished. Nevertheless, this case plays an important role in the interpretation of the nonlinear bifurcation scenario, below. The linear stability analysis of this case is very similar to that of the stratified case discussed before, except for the concentration gradient, which now reads (in dimensionless form) $\partial_z C = 1 - B/(\psi Ra)$, rather than $\partial_z C = - B/(\psi Ra)$ as before. This difference can be accounted for by substituting $B$ with $B - \psi Ra$ in the Eqs.(\ref{lam})-(\ref{omegac}) leading to a linear instability behavior qualitatively the same as before. The threshold of the stationary instability is now
\begin{equation}
Ra^{st,f}_c=\frac{B}{(L+1)\psi} \approx \frac{B}{\psi} \label{stfd}
\end{equation}
which is reduced by a factor of $1/L$ (at the same $B$) due to the destabilizing effect of thermodiffusion. On the other hand, sedimentation has to be stronger by that factor $L$, in order to be relevant. Thus, the oscillatory instability with the threshold
\begin{equation}
Ra^{os,f}_c=Ra^0+\frac{27 Pr}{14+27 Pr (\psi + 1)}(B-B_2) \label{onsfd}
\end{equation}
and the frequency at onset
\begin{equation}
\omega^2_{c,f} = \frac{6 Pr}{14+27 Pr(\psi + 1)} ({B} - B_2 ) \label{omegafd}
\end{equation}
occurs for $B > B_2 \equiv Ra^0 \psi$. The relevant sedimentation strength is now $B_2 \sim 10^4$. 
Again, the frequency decreases with increasing $Ra$ number and vanishes at a certain $Ra_{2,f}$ leading to a stationary instability. There is again a 'triple' point, $Ra^{st,f}_c(B_2) = Ra^{os,f}_c(B_2) = Ra_2(B_2)$ and for very large $B$, $Ra_2(B)  \propto ({B} - B_2)/(\psi+1)$, asymptotically. Eqs.\@ (\ref{stfd})--(\ref{omegafd}) correspond to Eqs.\@ (26) and (27) of Shliomis \& Smorodin [2005].

\bigskip

\noindent {\bf 4. Bifurcation Scenario} \smallskip

\noindent A linear theory can neither predict the actual geometry of the emerging pattern, nor can it give the its type and its saturation amplitude obtained in the long time limit. It also fails to describe the complete bifurcation topology, in particular it misses finite amplitude instabilities and sometimes it delivers (linear) instabilities that turn out to be transients, only. This happens for the thermal instability in colloidal suspensions (without sedimentation), in the case of a negative separation ratio $\psi$, where a linear oscillatory instability relaxes back to the quiescent stable state [Ryskin \& Pleiner 2005]. Since a similar linear oscillatory instability has been found in the previous section, it is compulsory in the present case to discuss the nonlinear behavior. This can be done most easily by numerical methods and will be described first, revealing in particular the time evolution into the saturation state. Afterwards, we present an approximate analytical expression of the amplitude of the (stationary) state that covers fairly well the important nonlinear features and allows for their understanding.   

We start with the same numerical method as already used in [Ryskin \& Pleiner 2005], section IV-A,  to investigate the time evolution of the system in the nonlinear regime. This method is essentially an extension of the Lorenz model [Lorenz 1963, Veronis 1965, Ahlers \& L\"ucke 1985]. The simplifying idea of this method is to look only for solutions corresponding to a 2-dimensional pattern (convection rolls). From the thermal instability behavior of colloidal binary mixtures one knows that the roll pattern is unstable against a square pattern (and for high $\psi$ values to certain kinds of cross-rolls) close to $Ra^0$ [Huke et al. 2007]. This might be still the case when sedimentation is included to the analysis, although no studies on that are available. Nevertheless, the basic notions of the nonlinear instabilities, which we will derive below, are certainly also true for square patterns, but much harder to get numerically and analytically than for the roll pattern.

For convection rolls, which are periodic with wave number $k$ in the lateral direction, the ansatz reads  
\begin{eqnarray}
C\left( {x,z,t} \right) &=&  C_0(z,t) + c_1(z,t) \cos{k x }, \label{conform}\\
T\left( {x,z,t} \right) &=&  - z +  \theta _0
\left( {z,t} \right) + \theta_1 \left( {z,t} \right)\cos{k x},
\label{Tform}\\
w (x,z,t) &=& w_1(z,t) \cos{k x}. \label{velozform}
\end{eqnarray}
taking into account five modes. The crucial difference to the previous work is that here we have an inhomogeneous boundary condition, Eq.\@ (\ref{bkc}), for the concentration variable. 

\begin{figure}
\includegraphics[width=8.6cm]{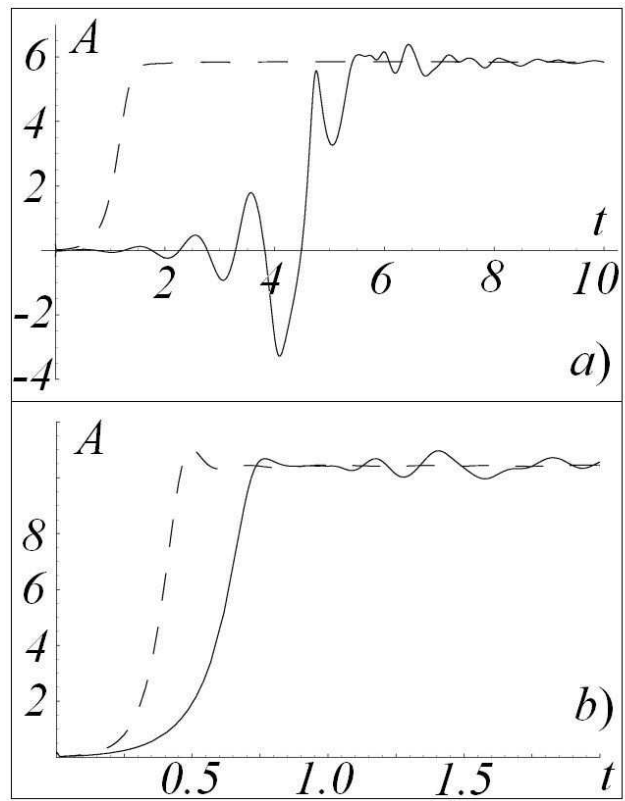}
\caption{The time evolution of the convection amplitude for $B=400$ at two different values of the Rayleigh number corresponding a) to the oscillatory linear regime, $Ra = 2200$, and b) to the stationary linear regime, $Ra=  3200$. The broken lines are appropriate reference solutions for the case without sedimentation ($B=0$). \label{fig1}}
\end{figure}

The initial state has zero velocity and a linear temperature profile, while for the concentration field we assume consecutively a homogeneous, a stratified, and a fully developed field, with $\partial_z C =0$, $\partial_z C = 1 - B/(\psi Ra)$, and $\partial_z C = - B/(\psi Ra)$, respectively, as discussed in the previous Section. We then apply additionally a small perturbation of the velocity field $w_1$ of the form $\sim \cos^2(\pi z)$. With typical parameter values, $\psi = 10$, $Pr = 7$, and $L = 10^{-4}$, and for either moderate $B = 400$ or rather high values $B=18000$ a $Ra$ number is chosen above the linear threshold value, thus monitoring the temporal development of the instabilities.  
In almost all cases we find that the system approaches a stationary state at long times, independent of the initial state (a rare exception is discussed below). However, how this stationary state is reached depends strongly on the linear behavior:  If for $B=400$ the $Ra$ number is chosen to be in the linear oscillatory regime, between $Ra_c^{os} < Ra <Ra_2$,  Eq.(\ref{rac}), the amplitude first starts to oscillate around zero, before it increases and, after some overshoot wiggles, reaches the asymptotic constant value, cf. Fig.\ref{fig1}a).  Choosing instead a $Ra$ number above $Ra_2$, where the system shows a stationary instability, linearly, the amplitude increases directly from zero to its final values, cf. Fig.\ref{fig1}b). This explains the reduction of the oscillation frequency with increasing Rayleigh number (Fig. 1 of Shliomis \& Smorodin [2005]) as a remnant of the linear behavior rather than a genuine nonlinear effect. In addition, we show that this transient frequency becomes zero, when the Rayleigh number exceeds $Ra_2$.

\begin{figure}
\includegraphics[width=8.6cm]{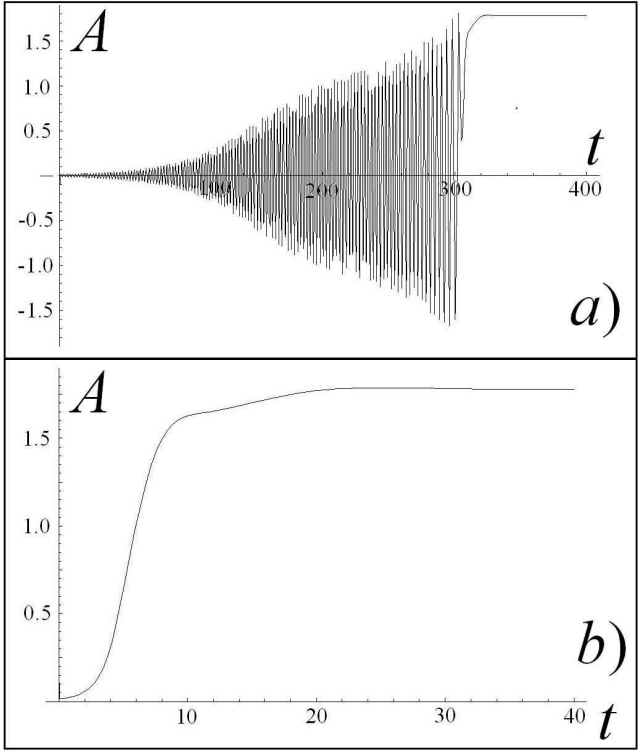} 
\caption{The amplitude of a finally steady convection as a function of time for different initial concentration profiles: a) the fully developed profile, and b) the homogenous concentration profile. Parameters are the same in both cases $Ra=1796$, $B=18000$, $\psi=10$, $L=10^{-4}$.   \label{fig2}}
\end{figure}

Similarly, and even more pronounced is this behavior for large values $B>B_2$ and a $Ra$ number above $Ra^{os,f}_c \approx Ra^0$: Starting from the homogeneous ground state, where the linear stability predicts a stationary instability, since $\psi$ is positive [Ryskin et al. 2003], the nonlinear numerical solution shows a smooth and monotonic transition to the final value of the amplitude, cf. Fig.\ref{fig2}b). Starting on the other hand from the fully developed ground state, where linear theory gives an oscillatory instability, Eqs.(\ref{onsfd}) and (\ref{omegafd}), the (nonlinear) amplitude oscillates for a rather long time around a zero value with growing peak amplitude, until finally it switches to the stationary finite value, cf. Fig.\ref{fig2}a). Starting from the stratified ground state, the system is linearly stable, since for such a large $B$ value $Ra_c^{os} \gg Ra^0$. However, with time the concentration profile evolves towards the fully developed one giving rise, at the end, to the behavior described before. The final state is the same stationary convection in all three cases. 
This behavior is in marked contrast to the case of the linear oscillatory instability when heating a system with negative separation ratio from below (without sedimentation), where the initial convective oscillations relaxes back to the \emph{non-convective} state [Ryskin \& Pleiner, 2005].

For the convection amplitude of the ultimate asymptotic stationary state we derive an approximate analytical expression in terms of $Ra$ and $B$.
Substituting Eqs.\@ (\ref{conform})-(\ref{velozform}) into the nonlinear equations of
motion (\ref{ns})-(\ref{conz}) and sorting out the different lateral
dependencies for the stationary state yields the following system of equations
\begin{eqnarray} \label{bv1} \left( {D^2  - k^2 } \right)^2 w_1 &=& Ra{\kern 1pt}
{\kern 1pt}
{\kern 1pt} k^2 ( {\theta_1  - \psi c_1}),\label{nsnl}\\
\frac{1}{2}\partial_z
\left( {w_1 c_1} \right)   &=& L \partial_z^2 ( C_0 +
\theta _0 ), \label{nlc1}\\
w_1 \partial_z C_0 &=& L
\left( {\partial_z^2  - k^2 } \right)( {c_1 + \theta_1 }),\label{C0n}\\
\frac{1}{2}
\partial_z \left( {w_1 \theta_1 } \right)  + w_1 &=& \partial_z^2 \theta_0,\label{thnl}\\
{\kern 1pt}
w_1 \partial_z \theta _0  &=& \left( {\partial_z^2  - k^2 }
\right)\theta_1,\label{T0nl}
\end{eqnarray}

Equation (\ref{nlc1}) can be integrated once. Taking into account the boundary condition Eq.\@ (\ref{bkc}) the concentration profile $c_1(z)$ is found to be
\begin{equation}
c_1=\frac{2L}{w_1}\left( \partial_z(C_0+\theta_0) - (1-\frac{B}{\psi Ra})\right) \label{c1n}
\end{equation}
Far from the boundaries $C_0$ and $c_1$ are proportional to $L$ . This follows from the requirement of Eq.\@ (\ref{C0n}) to be consistent with Eq.\@ (\ref{c1n}) and by taking into account that far from
the boundaries the derivatives of the functions are small. Thus,
in Eq.\@ (\ref{c1n}) $C_0$ can be neglected except close to
the boundaries and we get
\begin{equation}
c_1  =  - \frac{2 L}{w_1}\left(1 - \frac{B}{\psi Ra}  - \partial_z \theta_0\right) . \label{c1n2}\\
\end{equation}
To satisfy the boundary conditions for $c_1$, and to find the
profile of the concentration field near the boundaries, one needs
to solve the boundary layer problem. This has been done in Ryskin \& Pleiner [2004], Appendix A. It
was shown that the boundary layer depth $\delta \sim L^{1/3}$ is rather small and its contribution to the amplitude equation gives only small corrections $\sim L$. Therefore, Eq.\@ (\ref{c1n2}) can be used to find the velocity and temperature distributions.
This is obtained approximately by means of the trial functions 
\begin{equation}
w_1= A\cos^2(\pi z), \quad
\theta_0  = G \sin (2\pi z), 
\quad \theta_1 = F \cos \pi z
. \label{pf3}
\end{equation}
Substituting those profiles into Eqs.\@ (\ref{nsnl}),(\ref{thnl}), and (\ref{T0nl}) and projecting these equations onto the weight function $\cos^2(\pi z)$, leads to a
system of three algebraic equations for the amplitudes $A, F, G$. Solving for $A$, the saturation amplitude of convection, we find the implicit expression 
\begin{equation}
\left(1+ \frac{12}{5}\bar A\right)\left(Ra^0 \bar A - L \,[\psi Ra -B]\right) = Ra \bar A 
  \label{ampl}
\end{equation}
relating $\bar A \equiv A^2 / (32 \pi^2)$ to the driving force $Ra$, the material parameters $L$, $\psi$, and $B$.

\begin{figure}
\includegraphics[width=8.6cm]{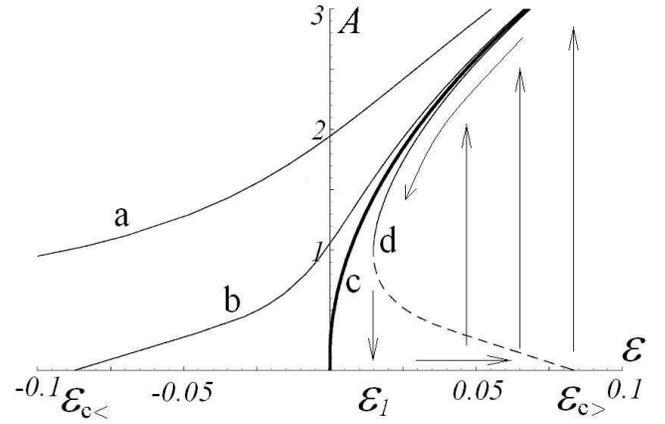} 
\caption{The amplitude of the stationary convection as a function of the reduced Rayleigh number $\varepsilon = (Ra-Ra^0)/Ra^0$ for different values of the barometric number  a) $B \to 0$, b) $B=16000$, c) $B=B_2\equiv Ra^0 \psi \approx 17534$, d) $B=19000$. Solid lines correspond to stable branches. The direction of the hysteresis loop is shown by the arrows.  \label{fig3}}
\end{figure}

In Fig.\@ \ref{fig3} the amplitude $A$ is shown as a function of the reduced Rayleigh number $\varepsilon  =(Ra-Ra^0)/Ra^0$, where $Ra^0$ is the linear threshold for thermal convection of a single component liquid. Two qualitatively different types of behavior are found depending on the barometric number $B$ as the crucial parameter. First, for $B \leq B_2 \equiv Ra^0 \psi$, there is a monotonic increase of the amplitude with the Rayleigh number to arbitrarily high $Ra$ numbers (lines a,b,c). These lines start with a vertical slope (not generally visible at the scale of the figure) at the threshold $\varepsilon_{c<}$, i.e. at the linear stability threshold $Ra^{st,f}_c = B/\psi$ (which is $ < Ra^0$ in this case). In the opposite case, $B>B_2$ (line d), there are two branches of the function $A(\varepsilon)$ in the interval $\varepsilon_1 < \varepsilon \leq \varepsilon_{c>}$. According to our numerical calculations the upper branch appears to be stable and the lower one unstable. Again, such lines (in particular their unstable branches) intersect the abscissa vertically at $\varepsilon_{c>}$, i.e. at the linear threshold $Ra^{st,f}_c$, which is now $ > Ra^0$. This is the scenario of a backward instability, where the amplitude takes a finite value $A_1$ at the threshold $\varepsilon_1$ with
\begin{eqnarray}
A_1^2 &=& \frac{40 \pi^2}{3} ( \sqrt{1+\varepsilon_1} - 1 )  \label{Aeps} \\
%
%
\varepsilon_1^2 &\approx&  \frac{16}{5} \,L \psi \, \beta \quad\quad {\rm for} \quad 
1> \beta > L \psi \label{eps1} \\
\varepsilon_1 &\approx& \beta \quad\quad\quad\quad\quad {\rm for} \quad \beta < L \psi
\label{eps2}
\end{eqnarray} 
where $\beta \equiv (B-B_2)/B_2$. Increasing the Rayleigh number beyond $\varepsilon_1$ the actual jump from a convection-free to the convective state takes place at a $Ra$ number that is generally higher than the $Ra$ number, at which the system jumps back from the convective to the convection-free state lowering $Ra$. This hysteretic behavior is indicated in Fig.\@ \ref{fig3} by arrows.

It may come as a surprise that for vanishing amplitudes thresholds are found, which belong to the linear stability analysis of the initially fully developed concentration profile (rather than the stratified or homogeneous one). The reason is that during the development of the convective patterns, the slow concentration dynamics also evolves until the final state is reached. Thus, the nonlinear behavior of Fig.\@ \ref{fig3} is obtained for any initial state, even the homogeneous one, but only after a long time. 

\begin{figure}
\includegraphics[width=8.6cm]{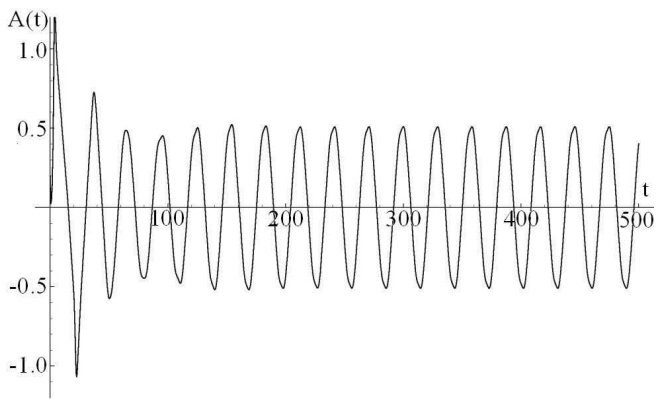}
\caption{The oscillatiory nonlinear amplitude as a function of time for $Ra=1755 \gtrsim Ra^0$, $B=17540 \gtrsim B_2$, $\psi=10$, and $L=10^{-4}$.    \label{fig4}}
\end{figure}

The last question to be discussed in this section concerns the existence of oscillatory, or more general non-stationary, nonlinear convective states. Numerically, in almost all cases a stationary instability has been found. Our approximative analytical solution, Eq.\@ (\ref{ampl}) shows that for $B<B_2$ there is always a stationary solution, while for $B>B_2$ the finite amplitude stationary state only exists for $\varepsilon \geqslant \varepsilon_1$. Of course, the convection-free state $A=0$ exists for all numbers $Ra$ and $B$, although it looses its stability at some $Ra_c(B)$, also depending on the initial state, as has been discussed by a linear analysis above. Now it can happen (only for $B>B_2$) that the convection-free state is already unstable with respect to a linear oscillatory instability, but no stationary nonlinear solution exists, leading to a situation where an oscillatory instability could be the natural response of the system. This requires that the threshold $Ra_c^{os,f}$, Eq.\@ (\ref{onsfd}), for an initially fully developed concentration profile, is lower than $\varepsilon_1$. This is possible in a very narrow parameter range only, where $\varepsilon_1$ is given by Eq.\@ (\ref{eps2}), i.e. for $B-B_2 < L \psi B_2$ and $Ra - Ra^0 < L \psi Ra^0$.
In this parameter range an oscillatory solution has indeed been found numerically, which is shown in Fig.\@ \ref{fig4}. 
The possibility to observe this and other solutions experimentally will be discussed in the following section.

\bigskip

\noindent {\bf 5. Conclusions} \smallskip

\noindent We have shown that effects of sedimentation significantly change the linear as well as the non-linear behavior of thermal convection in colloidal suspensions. We have considered the instability of three possible convection-free state - an initially homogenous one, an initially stratified one and the one with the fully developed concentration profile. In the two latter cases the linear instability can be oscillatory, if the strength of sedimentation, quantified by the barometric number, is sufficiently large. The nonlinear treatment, however, reveals that the oscillations are transient only, finally ending up in a stationary convective state, where, for simplicity, we have considered roll patterns only. Only in a very narrow window in the parameter space non-linear oscillations can exist. It is not obvious that it is possible to observe such a non-linear oscillatory convection in experiment, since in the numerics we had to tune the parameters $Ra$ and $B$ with an accuracy up to $0.1\%$ to get this state. On the other hand our 5-mode model of the nonlinear evolution is certainly not exact and we cannot guarantee that this numerical oscillatory solution would also be obtained in a more refined model. Finally, even if the oscillatory solution exists in a real physical system, it is certainly very difficult to tune the barometric number in experiments with an accuracy of $0.1\%$, since too many physical effects contribute to it.

In a certain range of the parameter space, in particular for very high barometric numbers, the  stationary convective solution comes in two branches, a stable and an unstable one. The former can coexist with the stable convection-free state, a situation that leads to a hysteretic behavior in experiments.
A hysteretic behavior was indeed observed in experiments [Bozhko \& Putin 2003]. Although a direct comparison with our theory is not possible, since the barometric number in the experiment  is not really known and most of the experimental results are obtained in the presence of a magnetic field, a situation which we have not considered so far. We hope that our investigations will motivate further experiments on sedimentation effects in the convective instabilities of colloidal suspensions.

\bigskip

\noindent {\bf References} \smallskip

\noindent $^*$ present address: Abtl. Theoretische Biologie, Universit\"at Bonn, 53115 Bonn, Germany; ryskin@uni-bonn.de

\noindent Ahlers, G. \& L\"ucke, M. [1987] ``Some properties of an eight-mode Lorenz model for convection in binary fluids'' {\it Phys. Rev.} {\bf A35}, 470.

\noindent Biben, T., Hansen, J.-P., \& Barrat, J.-L. [1993]
``Density profile of concentrated colloidal suspensions in sedimentation equilibrium'' {\it J. Chem. Phys.} {\bf 98}, 7330.

\noindent Blums, E., Mezulis, A., Maiorov, M., \& Kronkalns, G. [1997] ``Thermal diffusion of the magnetic nanoparticles in ferrocolloids: Experiment on particle separation in vertical columns'' {\it J. Magn. Magn. Mater.} {\bf 169},  220.

\noindent Blums, E., Odenbach, S., Mezulis, A., \& Maiorov, M. [1999] ``Magnetic Soret effect in hydrocarbon based colloid containing surfacted {\it Mn-Zn} ferrite particles'' {\it J. Magn. Magn. Mater.} {\bf 201},  268.

\noindent Blums, E. [2002]
``Heat and mass transfer phenomena'' {\it J. Magn. Magn. Mat.} {\bf 252}, 189.

\noindent Boussinesq, J. [1903]
{\it Th\'eorie Analytique de la Chaleur} (Gauthier-Villars, Paris) Vol.II,  p. 172. 

\noindent Bozhko, A. \& Putin, G. [2003] `Heat transfer and flow patterns in ferrofluid convection''
 {\it Magnetohydrodynamics} {\bf 30}, 147.

\noindent Bozhko, A., Putin, G., Beresneva, E., \& Bulychev, P. [2006] ``On magnetic field control experiments of ferrofluid convection motion'' {\it Z. Phys. Chem.} {\bf 220}, 1. 

\noindent Brand, H.R., Hohenberg, P.C.,  \& Steinberg, V. [1984]
``Codimension - 2 bifurcations for convection in binary mixtures'' {\it Phys. Rev.} {\bf A30}, 2548.

\noindent Cross, M.C. \& Hohenberg, P.C. [1993]  ``Pattern formation outside of equilibrium'' {\it Rev. Mod. Phys.} {\bf 49},  581.

\noindent Huke, B., L\"ucke, M., B\"uchel, P., \& Jung, C. [2000] ``Stability boundaries of roll and square convection in binary mixtures with positive separation ratio'' {\it J. of Fluid Mech.}  {\bf 408}, 121.

\noindent Huke, B., Pleiner, H., \& L\"ucke, M. 2007 ``Convection patterns in colloidal solutions" {\it Phys. Rev. E} {\bf 75}, 036203.

\noindent Kolodner, P., Williams, H., \& Moe, C. [1988] ``Optical measurement of the Soret coefficient of ethanol/water solutions`` {\it J. Chem. Phys.} {\bf 88}, 6512. 

\noindent Lenglet, J., Bourdon, A., Bacri, J.-C., \& Demouchy, G.  [2002] ``Thermodiffusion in magnetic colloids evidenced and studied by forced Rayleigh scattering'' {\it Phys. Rev.} {\bf E65}, 031408. 

\noindent Lorenz, E. N. [1963] ``Deterministic non-periodic flow'' {\it J. Atmos. Sci.} {\bf 20}, 130. 

\noindent L\"{u}cke, M., Barten, W., B\"{u}chel, P., F\"{u}tterer, C., 
Hollinger, St., \& Jung, Ch. [1998] ``Pattern formation in binary fluid
convection and in system with throughflow" in {\it Evolution of Spontaneous
Structures in Continuous Systems} (eds. Busse, F. H. \& 
M\"{u}ller, S. C.), Lecture Notes in Physics Vol. {\bf 55} (Springer, Berlin) 127.

\noindent {Milner, S.T.} [1989] ``Hydrodynamics of semidilute polymer solutions'' {\it Phys. Rev. Lett.} {\bf 66}, 1477. 

\noindent {Milner, S.T.} [1993] ``Dynamical theory of concentration fluctuations in polymer solutions under shear''  {\it Phys. Rev.} {\bf E48}, 3674.

\noindent {Onuki, A. }[1989]  ``Elastic effects in the phase transition of polymer solutions under shear flow'' {\it Phys. Rev. Lett.} {\bf  62}, 2472.

\noindent Platten, J.K. \& Chavepeyer, G. [1976]
``Instabilit\'e et flux de chaleur dans le probl\'eme de B\'enard \'a deux constituants aux coefficients de Soret positifs'' {\it Int. J. Heat Mass Transf.} {\bf 19}, 27.

\noindent Platten, J.K. \& Legros, J.C. [1984] {\it Convection in Liquids} (Springer, Berlin).

\noindent Pleiner, H. \& Harden, J.L., [2003] ``General 2-Fluid Hydrodynamics of Complex Fluids and Soft Matter", in {\it Nonlinear Problems of Continuum Mechanics,  Special issue of  Notices of Universities. South of Russia. Natural sciences},  p.46; also available from arXive cond-mat/0404134.

\noindent Rosensweig, R.E. [1985] {\it Ferrohydrodynamics} (Cambridge University Press, Cambridge, England).

\noindent Ryskin, A., M{\"u}ller, H.-W., \& Pleiner, H. [2003] ''Thermal convection in binary mixtures with a weak concentration diffusivity, but strong solutal buoyancy forces`` {\it Phys. Rev.} {\bf E67}, 046302.

\noindent Ryskin, A. \& Pleiner, H.  [2004] ''The influence of a magnetic field on the Soret-dominated thermal convection in ferrofluids`` {\it Phys. Rev.} {\bf E69}, 046301.

\noindent Ryskin, A. \& Pleiner, H. [2005] ''Thermal convection in colloidal suspensions with negative separation ratio`` {\it Phys. Rev.} {\bf E71}, 056303.

\noindent Ryskin, A. \& Pleiner, H. [2007] ``Magnetic field driven instability in stratified ferrofluids`` {\it Phys. Rev.}, {\bf E75}, 056303.

\noindent Shliomis, M.I. \& Smorodin, B.L. [2005]
``Onset of convection in colloids stratified by gravity'' {\it Phys. Rev.} {\bf E71}, 036312.

\noindent Tynj{\"a}l{\"a}, T., Bozhko, A., Bulychev, P., Putin, G., \& Sarkomaa, P. [2006] ``On features of ferrofluid convection caused by barometrical sedimentation" {\it J. Magn. Magn. Mater.} {\bf 300}, 195.

\noindent Veronis, G. [1965] ``On finite amplitude instability in thermohaline convection'' {\it J. Mar. Res.} {\bf 23}, 1.

\noindent Winkler, B.L. \& Kolodner, P. [1992]
``Measurements of the concentration field in nonlinear traveling-wave convection'' {\it J. Fluid Mech.}   {\bf 240} 31. 
\end{document}